\documentclass[amsmath,amsfonts,aps,nofootinbib,preprintnumbers]{revtex4}
\usepackage{graphicx}
\usepackage{amssymb}
\usepackage{epsfig}
\begin{document}
\baselineskip=20pt
\title{The nucleon magnetic moment in the $\epsilon$-regime of HBChPT}
\author{Joseph Wasem}\email{wasem@u.washington.edu} \affiliation{Department of Physics,
  University of Washington\\ Box 351560, Seattle, WA 98195, USA}

\preprint{NT@UW-08-02}

\begin{abstract}
The nucleon magnetic moment is calculated in the $\epsilon$-regime
of Heavy Baryon Chiral Perturbation Theory to order $\epsilon^3$,
using the method of collective variables to integrate
nonperturbative pion zero modes. Contributions containing multiple
sources of zero modes enter, allowing for charge-carrying zero mode
pion fields that connect the sources. The result of this calculation
will allow for lattice QCD calculations involving nucleons to
systematically extract the leading low energy coefficients of Heavy
Baryon Chiral Perturbation Theory with electromagnetic interactions.
\end{abstract}

\maketitle

\section{Introduction}
As the only known way of computing nuclear observables directly from
QCD, lattice QCD has received steadily increasing attention in
recent years. However, limited computational resources have required
calculations to use unphysically large quark masses and severely
restricted computational volumes. Some recent calculations have used
pion masses that are still a factor of two greater than the physical
value, with a spatial lattice volume no greater than a few Fermis on
a side. Both of these restrictions necessitate the use of Chiral
Perturbation Theory (ChPT) in finite volume to extrapolate the large
mass, finite volume lattice results to physical mass, infinite
volume results that can be directly compared to experiment.

Heavy Baryon Chiral Perturbation Theory
(HBChPT)\cite{Jenkins:1991ne,Jenkins:1990jv} is used when
calculating baryon properties. When electromagnetism is included,
simple extensions to the theory that maintain the chiral symmetry
are needed and are given in refs.
\cite{Coleman:1961jn,Gasser:1983yg,Gasser:1984gg,Jenkins:1992pi}.
Typically the small expansion parameters in HBChPT (in what is known
as the $p$-regime) are $p/m_{B}$, $p/\Lambda_{\chi}$, and
$m_{\pi}/\Lambda_{\chi}$ where $p$ is the typical momentum and
$\Lambda_{\chi}$ is the chiral symmetry breaking scale, typically of
order 1 GeV\cite{Luscher:1985dn}. However, as the zero momentum pion
propagator is given by $1/m_{\pi}^2 V$ (where $V$ is the spacetime
volume), for small quark masses the pion zero momentum modes will
become enhanced relative to nonzero modes and become nonperturbative
near the chiral limit\cite{Gasser:1987ah}. The regime where this
occurs is known as the $\epsilon$-regime, and in this regime a
different counting scheme is used which accounts for the
nonperturbative zero
modes\cite{Gasser:1987ah,Hansen:1990un,Hansen:1990yg,Hansen:1990kv,
Hasenfratz:1989pk,Hasenfratz:1989ux,Leutwyler:1992yt}. In the
$\epsilon$-regime the counting parameter is defined by $\epsilon
\sim 2\pi / \Lambda_{\chi}L \sim 2\pi / \Lambda_{\chi} \beta$ and
$\epsilon^2 \sim m_{\pi} / \Lambda_{\chi}$ where $L$ and $\beta$
are, respectively, the spatial and temporal dimensions of the box
one is performing the calculation in and $m_{\pi}$ is the pion mass.
The inclusion of decuplet baryons in the calculation further
requires the definition $\epsilon^2 \sim \Delta / \Lambda_{\chi}$,
with $\Delta$ the decuplet mass splitting. The nonperturbative zero
modes are integrated over using the method of collective
variables\cite{Gasser:1987ah} and baryons are included in the
collective variable framework using the method from ref.
\cite{Smigielski:2007pe}.

The magnetic moment has been the subject of intense study by the
lattice QCD community, including theoretical studies in
finite\cite{Beane:2004tw,Detmold:2004ap,Tiburzi:2007ep} and infinite
volume\cite{Caldi:1974ta,Jenkins:1992pi,Meissner:1997hn,Durand:1997ya,Hemmert:2002uh}
as well as lattice
calculations\cite{Leinweber:1990dv,Draper:1991uu,Gadiyak:2002wc,Gockeler:2003ay,Gockeler:2006ui,Gockeler:2007ir,Gockeler:2007hj,Alexandrou:2006ru}.
Most of these have involved the $p$-regime. With faster computers
becoming available, lattice theorists will be faced with a choice:
simultaneously decrease quark masses and increase lattice sizes so
as to remain in the $p$-regime, or keep lattice sizes constant and
push to lower and more physical quark masses leading directly into
the $\epsilon$-regime of ChPT. To facilitate the analysis of data
from lattice calculations in the $\epsilon$-regime this work focuses
on the calculation of the magnetic moment of the nucleons in HBChPT
in the $\epsilon$-regime to order $\epsilon^3$, with exact
integration of the pion zero momentum modes using the method from
ref. \cite{Smigielski:2007pe}.

\section{Heavy Baryon Lagrangian with Electromagnetism}
The low energy HBChPT Lagrangian that is consistent with
spontaneously broken $SU(2)_{L}\otimes SU(2)_{R}$ is, at leading
order\cite{Jenkins:1990jv,Jenkins:1991ne}:
\begin{eqnarray}\label{lagrangian}
    \mathcal{L}_0&=&\bar{N}iv\cdot\mathcal{D}N-\bar{T}_{\mu}iv\cdot\mathcal{D}T^{\mu}+\Delta\bar{T}_{\mu}T^{\mu}
    +\frac{f^2}{8}{ \rm Tr}[\partial_{\mu}\Sigma^{\dagger}\partial^{\mu}\Sigma]+\lambda\frac{f^2}{4}{ \rm Tr}[m_{q}\Sigma^{\dagger}+h.c.]\nonumber\\
    &&+2g_{A}^{0}\bar{N}S^{\mu}\mathcal{A}_{\mu}N
    +g_{\Delta N}[\bar{T}^{abc,\nu}\mathcal{A}^{d}_{a,\nu}N_{b}\epsilon_{cd}+h.c.]+2g_{\Delta\Delta}\bar{T}_{\nu}S^{\mu}\mathcal{A}_{\mu}T^{\nu}
\end{eqnarray}
with the nucleon fields $N$, the Rarita-Schwinger fields $T^{\mu}$
describing the $\Delta$-resonances, and the definitions:
\begin{eqnarray}
    \Sigma&=&\xi^{2}={\rm exp}\left(\frac{2iM}{f}\right),\nonumber\\
    M&=&\left(\begin{matrix}
    \pi^0/\sqrt{2} & \pi^+\cr \pi^- &
    -\pi^0/\sqrt{2}
    \end{matrix}\right), \nonumber\\
    \mathcal{A}^{\mu}&=&\frac{i}{2}(\xi\partial^{\mu}\xi^{\dagger}-\xi^{\dagger}\partial^{\mu}\xi),\nonumber\\
    V^{\mu}&=&\frac{1}{2}(\xi\partial^{\mu}\xi^{\dagger}+\xi^{\dagger}\partial^{\mu}\xi),\nonumber\\
    \mathcal{D}^{\mu}&=&\partial^{\mu}+V^{\mu}.
\end{eqnarray}
where the constant $f=132$ MeV, $\Delta$ is the decuplet mass
splitting, and the matrix $m_q$ is the quark mass matrix. The pion
fields are encapsulated in the matrix $M$. The Rarita-Schwinger
fields contain the $\Delta$-resonances according to:
\begin{equation}
    \begin{matrix}
    T^{111}=\Delta^{++}, & T^{112}=\frac{1}{\sqrt{3}}\Delta^{+}, & T^{122}=\frac{1}{\sqrt{3}}\Delta^{0},
    & T^{222}=\Delta^{-}
    \end{matrix}
\end{equation}
while the nucleons are an SU(2) vector given by
\begin{equation}
    N=\left(\begin{matrix}
    p\cr n \end{matrix}\right).
\end{equation}
The three couplings given in eqn. (\ref{lagrangian}) ($g_{A}^{0}$,
$g_{\Delta N}$, and $g_{\Delta\Delta}$) are the infinite volume,
chiral limit couplings between baryons and pions. $S^{\mu}$ is the
covariant spin vector and $v^{\mu}$ is the heavy baryon four
velocity with $v^2=1$ (typically calculations are done in the baryon
rest frame with $v^{\mu}=(1,\vec{0})$).

To calculate the magnetic moment in HBChPT, nucleon and decuplet
magnetic moment operators are necessary along with electromagnetic
couplings. Specifically, Feynman graphs that have a component
proportional to either $\bar{N}F^{\mu\nu}\sigma_{\mu\nu}N$ or
$\bar{N}F^{\mu\nu}\sigma_{\mu\nu}\tau^{3}N$ are sought. The lowest
dimension operators that achieve this are given by the
Coleman-Glashow SU(3) relations (adapted to two flavor HBChPT)
\cite{Coleman:1961jn,Jenkins:1992pi,Detmold:2004ap}:
\begin{eqnarray}
    \mathcal{L}_{mag1}&=&\frac{e}{4m_B}F_{\mu\nu}\left(\mu_{D}{\rm{Tr}}\left(\bar{\psi}\sigma^{\mu\nu}\psi\right)
    +\mu_{F}{\rm{Tr}}\left(\bar{\psi}\sigma^{\mu\nu}\tau^{3}_{\xi+}\psi\right)\right)\nonumber\\
    &\to&\frac{e}{4m_B}F_{\mu\nu}\left(\mu_{0}\bar{N}\sigma^{\mu\nu}N+\mu_{1}\bar{N}\sigma^{\mu\nu}\tau^{3}_{\xi+}N\right)
\end{eqnarray}
where the $\psi$ are octet baryon spinors and
$\tau^{a}_{\xi+}=\frac{1}{2}(\xi^{\dagger}\tau^{a}\xi+\xi\tau^{a}\xi^{\dagger})$.
The decuplet and nucleon-decuplet transition magnetic moment
operators are determined by finding SU(3) invariants as in ref.
\cite{Jenkins:1992pi} and adapting these to the two flavor theory.
These contributions give the additional terms in the
Lagrangian\cite{Jenkins:1992pi}
\begin{eqnarray}\label{magmomlagrangian}
    \mathcal{L}_{mag2}&=&-i\frac{3e}{m_{B}}\mu_{C}\bar{T}^{\mu,abc}Q^{d}_{a}T^{\nu}_{bcd}F_{\mu\nu}
    +i\frac{e}{2m_{B}}\mu_{T}F_{\mu\nu}\left(\bar{N}^{a}S^{\mu}Q^{b}_{a}T^{\nu}_{bcd}\epsilon^{cd}
    +\bar{T}^{\mu \ abc}S^{\nu}Q^{d}_{a}N_{b}\epsilon_{cd}\right).
\end{eqnarray}
Electromagnetism is included into HBChPT by substituting in eqn.
(\ref{lagrangian}) for $V^{\mu}$, $A^{\mu}$, and
$\partial_{\mu}\Sigma$ according to\cite{Jenkins:1992pi}:
\begin{eqnarray}
    V^{\mu}&\longrightarrow&V^{\mu}+\frac{1}{2}ie\mathcal{A}^{\mu}\left(\xi^{\dagger}Q\xi
    +\xi Q\xi^{\dagger}\right) \nonumber\\
    A^{\mu}&\longrightarrow&A^{\mu}-\frac{1}{2}ie\mathcal{A}^{\mu}\left(\xi Q\xi^{\dagger}
    -\xi^{\dagger}Q\xi\right) \nonumber\\
    \partial^{\mu}\Sigma&\longrightarrow&\mathcal{D}^{\mu}\Sigma=\partial^{\mu}\Sigma
    +ie\mathcal{A}^{\mu}\left[Q,\Sigma\right]
\end{eqnarray}
and in eqn. (\ref{magmomlagrangian}) for $Q$ with:
\begin{eqnarray}
    Q&\longrightarrow&\frac{1}{2}(\xi{Q}\xi^{\dagger}+\xi^{\dagger}Q\xi).
\end{eqnarray}

Higher order Lagrangian terms that will also be important include:
\begin{eqnarray}\label{1/m lagrangian}
    \mathcal{L}_{1}&=&-\left(\bar{N}\frac{\mathcal{D}^2-(v\cdot\mathcal{D})^2}{2m_B}N\right)
    +\left(\bar{T}^{\mu}\frac{\mathcal{D}^2-(v\cdot\mathcal{D})^2}{2m_B}T_{\mu}\right)\nonumber\\
    &\longrightarrow&\bar{N}\frac{\vec{\partial}^2}{2m_B}N-\bar{T}^{\mu}\frac{\vec{\partial}^2}{2m_B}T_{\mu}.
\end{eqnarray}
The chirally covariant derivatives $\mathcal{D}$ contain $V^{\mu}$
terms which will not be important at the order considered and so are
dropped. The coefficients of these terms can be determined by
reparametrization invariance\cite{Luke:1992cs} or by matching to the
relativistic theory in the path integral\cite{Bernard:1992qa}. The
resulting terms in eqn. (\ref{1/m lagrangian}) can be included in
the baryon and decuplet propagators. Following an expansion in
$1/m_B$ one will then recover the terms equivalent to an insertion
of this kinetic energy operator. Other $1/m_B$ operators exist
beyond these kinetic terms, but they will not enter until order
$\epsilon^4$ or higher, and so are not important for this
calculation.

\section{Zero Mode Integration and Results}
In calculating the nucleon magnetic moment in the $\epsilon$-regime
the pion zero modes will be determined nonperturbatively using the
method described in ref. \cite{Smigielski:2007pe}. Many of the
graphs contributing to the magnetic moment will have a zero mode
structure similar to that of the graphs in ref.
\cite{Smigielski:2007pe}, thus simplifying the calculation. In
making the change of variables to allow the zero mode integration, a
Jacobian factor arises which adds another term (of order
$\epsilon^2$) to the Lagrangian\cite{Hansen:1990un}:
\begin{equation}
    \mathcal{L}_{Jacobian}=\frac{8}{3f^2\beta L^3}\left(\frac{1}{2}\hat{\pi}_{0}^{2}+\hat{\pi}_{+}\hat{\pi}_{-}\right)
\end{equation}
where the $\hat{\pi}$ fields are nonzero mode pion fields (following
the conventions of ref. \cite{Smigielski:2007pe}). This has the form
of a shift in the squared pion mass of
$\Delta{m_{\pi}^2}=8/3f^2\beta{L^3}$. When applied to the pion
propagators in the order $\epsilon$ graphs the contribution comes in
at order $\epsilon^3$, however because it has the form of a mass
correction it will be placed in that position.

\begin{figure}[!h]
\centering
\includegraphics[scale=.75]{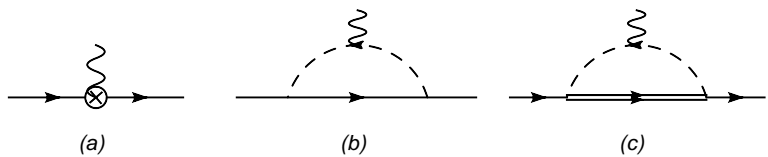}
\caption{(a)Leading and (b,c)next to leading order contributions to
the nucleon magnetic moment.} \label{LOgraphs}
\end{figure}

\begin{figure}[!h]
\centering
\includegraphics[scale=.75]{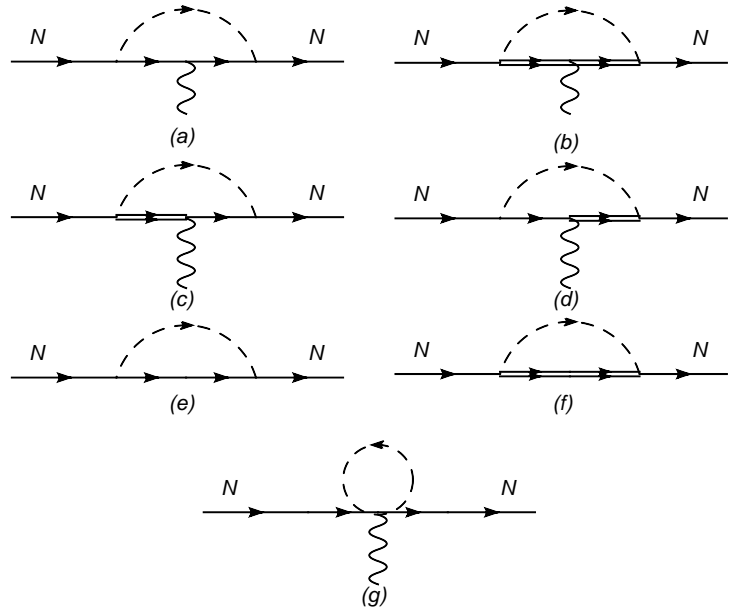}
\caption{Contributions to the magnetic moment at order $\epsilon^2$.
Graphs (e) and (f) are wavefunction renormalization diagrams.}
\label{e2graphs}
\end{figure}

The leading and next to leading order contributions to the nucleon
magnetic moment are given in fig. (\ref{LOgraphs}).
The contribution from fig. (\ref{LOgraphs}a) stems from two
different Lagrangian terms in eqn. (\ref{magmomlagrangian}), one
proportional to $\mu_0$ and one proportional to $\mu_1$. As the term
proportional to $\mu_0$ does not contain any pion field operators it
cannot have any zero mode structure, while the term proportional to
$\mu_1$ has zero mode structure similar to an insertion of the axial
current in ref. \cite{Smigielski:2007pe}. The two order $\epsilon$
graphs will have zero mode contributions from the electromagnetic
vertex only. At order $\epsilon^2$ the graphs in fig.
(\ref{e2graphs}) are found to contribute, where figs.
(\ref{e2graphs}e) and (\ref{e2graphs}f) are the leading wavefunction
renormalization contributions. Again these contributions are very
similar to those found in the calculation of the axial charge, and
as such have the same zero mode structure.

\begin{figure}[!h]
\centering
\includegraphics[scale=.75]{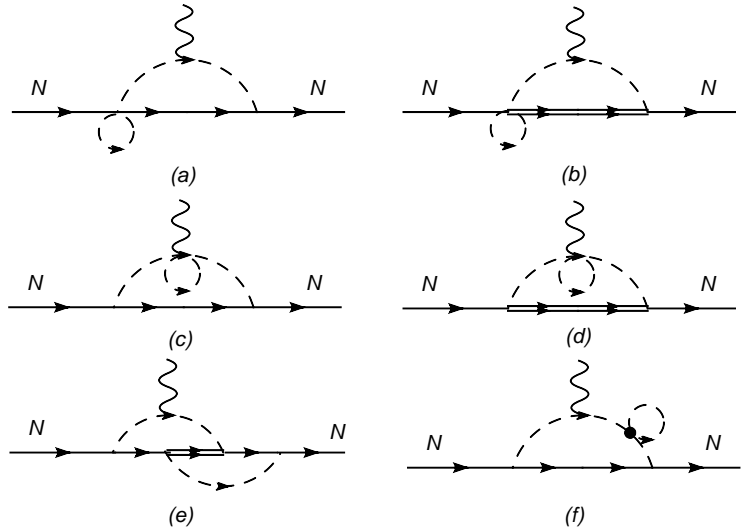}
\caption{Examples of contributions to the magnetic moment at order
$\epsilon^3$.} \label{e3graphs}
\end{figure}

At order $\epsilon^3$ most of the new contributing diagrams are
formed by attaching a simple pion loop at every vertex in an order
$\epsilon$ diagram, as in fig. (\ref{e3graphs}a-d). In some cases
(figs. (\ref{e3graphs}c) and (\ref{e3graphs}d)) this does not create
any new zero mode structure, as all of the zero mode information
comes from the electromagnetic vertex. However, in fig.
(\ref{e3graphs}a) and (\ref{e3graphs}b) the addition of the pion
loop to the nucleon-pion vertex creates a new source of zero modes
in addition to those from the electromagnetic vertex. The zero mode
functions from contributions of this form separate into two groups
of diagrams. The first group (characterized by the function $A(s)$
as defined in the appendix) contains intermediate nonzero mode pions
where charge is conserved separately within both the zero mode and
nonzero mode sectors of the theory. The second group (characterized
by the function $B(s)$) contains intermediate zero mode pions which
carry electric charge from one vertex to the other. The last two
diagrams in fig. (\ref{e3graphs}) are not as simply constructed, but
are straightforward to compute. Figure (\ref{e3graphs}e) is the
lowest order two loop construction (the electromagnetic vertex can
go on either pion loop with an equal contribution) while fig.
(\ref{e3graphs}f) is an insertion of the pion kinetic energy
operator into an order $\epsilon$ diagram.

Once the various zero mode functions have been determined, the
remaining finite volume nonzero mode Feynman diagrams can be written
in the standard way as a four component sum over integers with the
zero vector specifically removed. Using the Abel-Plana formula, the
fourth component of this vector sum can be written as an infinite
volume integral plus finite volume corrections. As shown in ref.
\cite{Smigielski:2007pe}, due to the nature of the HBChPT
propagators, these correction terms will be negligible until at
least order $\epsilon^6$ and so can be ignored. Using this fact, the
diagrams can be determined by integrating over the time component of
the momentum and then summing over the remaining spatial components.
To perform the sums that arise, one can explicitly expand them in
powers of $\epsilon$ to the order desired, resulting in purely
numerical sums (expressed below as coefficients defined in the
appendix).

Calculating the contribution from each of the graphs given in figs.
(\ref{LOgraphs}-\ref{e3graphs}) results in:
\begin{eqnarray}
    \hat{\mu}&=&\hat{\mu}_0+\hat{\mu}_1+\hat{\mu}_2+\hat{\mu}_3\\
    \hat{\mu}_{0}&=&\mu_{0}+\mu_{1}\tau^{3}Q(s)\\
    \hat{\mu}_{1}&=&\frac{8m_{B}(g_{A}^{0})^2}{3f^2}\tau^{3}Q(s)
    \left(\frac{c_2}{8\pi^2L}-\frac{c_4L(m_{\pi}^2(s)+8/3f^2\beta L^3)}{16\pi^4}+\frac{1}{2m_{B}}\frac{3c_1}{8\pi L^2}+\frac{1}{4m_{B}^2}\frac{c_0}{L^3}
    \right)+\frac{16m_{B}g_{\Delta
    N}^{2}}{27f^{2}}\tau^{3}Q(s)
    \left(\frac{c_2}{8\pi^2L}\right.\nonumber\\
    &&\left.-\frac{3\Delta c_3}{32\pi^3}
    +\frac{(\Delta^2-(m_{\pi}^2(s)+8/3f^2\beta
    L^3))Lc_4}{16\pi^4}
    +\frac{1}{2m_{B}}\left(\frac{3c_1}{8\pi L^2}
    -\frac{\Delta c_2}{2\pi^2L}\right)+\frac{1}{4m_{B}^2}\frac{c_0}{L^3}\right)
\end{eqnarray}
\begin{eqnarray}
    \hat{\mu}_{2}&=&-\frac{m_{B}(g_{A}^{0})^2}{2{\pi}f^2}\tau^{3}Q(s)\sqrt{m_{\pi}^2(s)+\frac{8}{3f^2\beta L^3}}
    -\frac{4m_{B}g_{\Delta
    N}^{2}}{9{\pi}f^{2}}\tau^{3}Q(s)\mathcal{F}_{\pi}\nonumber\\
    &&-\frac{(g_{A}^{0})^2}{6f^2}\left(3\mu_{0}-\mu_{1}\tau^{3}Q(s)\right)
    \left(\frac{c_1}{4\pi
    L^2}+\frac{1}{m_{B}}\frac{c_0}{2L^3}\right)
    +\frac{20g_{\Delta
    N}^2}{27f^2}\mu_{C}\left(1+\frac{10}{3}\tau^{3}Q(s)\right)
    \left(\frac{c_1}{4\pi L^2}-\frac{\Delta c_2}{4\pi^2
    L}+\frac{1}{m_{B}}\frac{c_0}{2L^3}\right)\nonumber\\
    &&+\frac{16}{27}\frac{g_{\Delta
    N}g_{A}^{0}}{f^2}\tau^{3}Q(s)\mu_{T}\left(\frac{c_1}{4\pi L^2}-\frac{\Delta c_2}{8\pi^2
    L}+\frac{1}{2m_B}\frac{c_0}{L^3}\right)-\mu_{1}T(s)\tau^{3}\frac{c_1}{2\pi f^2 L^2}\nonumber\\
    &&-\left(\mu_{0}+\mu_{1}\tau^{3}Q(s)\right)\left(\frac{3(g_{A}^{0})^{2}}{2f^2}
    \left(\frac{c_1}{4\pi L^2}+\frac{c_0}{2m_{B}L^3}\right)
    +\frac{4}{3}\frac{g_{\Delta
    N}^2}{f^2}\left(\frac{c_1}{4\pi L^2}-\frac{\Delta c_2}{4\pi^2
    L}+\frac{c_0}{2m_{B}L^3}\right)\right)
\end{eqnarray}
\begin{eqnarray}
    \hat{\mu}_{3}&=&\frac{c_1c_2m_B}{6\pi^3f^4L^3}\tau^{3}\left(-\frac{2}{3}A(s)-\frac{1}{6}B(s)\right)\left(g_{A}^{0}\right)^{2}
    +\frac{2c_1c_2m_B}{81\pi^3f^4L^3}\tau^{3}\left(-A(s)+B(s)\right)g_{\Delta
    N}^{2}\nonumber\\
    &&+\frac{c_1c_2(g_{A}^{0})^2m_B}{6\pi^3f^4L^3}\tau^{3}Q(s)+\frac{c_1c_2g_{\Delta
    N}^2m_B}{27\pi^3f^4L^3}\tau^{3}Q(s)\nonumber\\
    &&-\frac{c_1c_2m_B}{8f^4\pi^3L^3}\left((g_{A}^{0})^4+\frac{16g_{\Delta
    N}^{4}}{81}+\frac{10}{9}g_{\Delta
    N}^2(g_{A}^{0})^{2}\right)\tau^{3}Q(s)-\frac{2c_1c_2(g_{A}^{0})^2m_B}{9\pi^3f^4L^3}\tau^{3}Q(s)
    -\frac{2c_1c_2g_{\Delta
    N}^{2}m_B}{81\pi^3f^{4}L^3}\tau^{3}Q(s)\nonumber\\
    &&+\frac{c'_3m_B}{36\pi^3f^4L^3}\tau^{3}Q(s)\left((g_{A}^{0})^{4}+\frac{16}{3}(g_{A}^{0})^{2}g_{\Delta
    N}^{2}-\frac{8}{81}g_{\Delta N}^{4}-\frac{56}{81}g_{A}^{0}g_{\Delta\Delta}g_{\Delta N}^{2}-\frac{400}{729}g_{\Delta
    N}^{2}g_{\Delta\Delta}^{2}\right)
\end{eqnarray}
with
\begin{equation}
    \pi\mathcal{F}_{\pi}=\sqrt{\Delta^2-\left(m_{\pi}^2(s)+\frac{8}{3f^2\beta L^3}\right)}{\rm
    log}\left(\frac{\Delta-\sqrt{\Delta^2-(m_{\pi}^2(s)+8/3f^2\beta L^3)}}{\Delta+\sqrt{\Delta^2-(m_{\pi}^2(s)+8/3f^2\beta L^3)}}\right)-\Delta{\rm
    log}\left(\frac{m_{\pi}^2(s)+8/3f^2\beta L^3}{\mu^2}\right)
\end{equation}
where the subscript refers to the order in $\epsilon$ of the
contribution. The zero mode functions $Q(s)$, $T(s)$, $A(s)$, and
$B(s)$ as well as the coefficients $c_0$, $c_1$, $c_2$, $c_3$,
$c_4$, and $c'_3$ are defined in the appendix. The pion mass also
has a zero mode contribution that has been previously
calculated\cite{Gasser:1987ah} which effects the replacement
$m_{\pi}^2\to m_{\pi}^2(s)$. The zero mode structure of this
replacement is the same as that from the tadpole diagram in fig.
(\ref{e2graphs}g) and indeed $m_{\pi}^2(s)=m_{\pi}^2T(s)$. Note that
the infinite volume portions of the order $\epsilon$ contributions
enter at order $\epsilon^2$, and are the first two terms in
$\hat{\mu}_{2}$.

\section{Discussion and Conclusion}
\begin{figure}[!h]
\centering
\includegraphics[scale=.75]{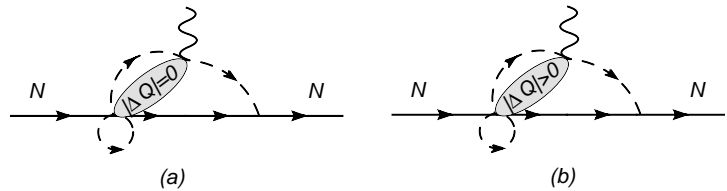}
\caption{Examples of the type of zero mode pions that contribute to
the functions (a)$A(s)$ with charge neutral zero mode configurations
(grey blobs) connecting each vertex and (b)$B(s)$ with charge
carrying zero mode configurations connecting each vertex.}
\label{zeroexamples}
\end{figure}

We have computed the nucleon magnetic moment in the $\epsilon$
regime to order $\epsilon^3$. In doing so, two new zero mode
functions have been calculated, $A(s)$ and $B(s)$. Both of these
functions have intriguing features, as each stem from diagrams with
two separate vertices that produce zero modes. The zero mode effects
contained in the function $A(s)$ are generated from zero mode fields
that connect the vertices but do not carry charge, as shown in fig.
(\ref{zeroexamples}a). This is in contrast to the zero modes
encapsulated in the function $B(s)$, as they connect and carry
charge between the two vertices, as shown in fig.
(\ref{zeroexamples}b). When writing down contributing diagrams where
the zero modes have been removed and placed in the function $B(s)$
the diagrams will appear to violate local charge conservation. In
fact, charge is locally conserved when the zero modes are included.
The form of the function $B(s)$, as shown in fig. (\ref{zerofunct}),
is easy to understand. As $s\to\infty$ the zero mode fluctuations
become damped and contributions of this type disappear due to charge
conservation in the nonzero mode pions, while as $s\to0$ the charged
zero modes will become increasingly important.

The behavior of the zero mode function $A(s)$ is unique because it
changes sign as $s\to0$, raising an interesting possibility. For a
given choice of spatial volume and pion mass, the temporal dimension
of a lattice calculation could be chosen such that $A(s)$ was zero,
thus removing these contributions from the calculation. In addition,
previous studies (see refs.
\cite{Detmold:2004ap,Detmold:2004qn,Li:2003jn}) have discussed the
changes that occur in the coefficients $c_1$ and $c_2$ in asymmetric
spatial volumes. Specifically, these coefficients can be either
enhanced or made to vanish with the correct choice of spatial
dimensions. If one was to combine these spatial dimension effects on
the coefficients with the temporal dimension effects on the zero
mode functions one could design a lattice with a significant degree
of control over the contribution of certain graphs.

Due to the increased complexity of lattice calculations involving
electromagnetism over lattice calculations with pure QCD, the
lattice theorist may be tempted to forgo working on quantities such
as the magnetic moment in favor of quantities such as the axial
charge. However, the electromagnetic properties of the nucleons are
some of the most precisely measured quantities in physics, and as
such an important test of lattice QCD will be the reproduction of
these values. Because of this, lattice calculations of the nucleon
magnetic moment in the $\epsilon$-regime will be of importance in
the coming years. Furthermore, configurations used to calculate
quantities such as the axial charge can also be used to calculate
the magnetic moment. In fact, calculating both quantities
simultaneously and performing the appropriate correlated statistical
analysis on each would be a powerful tool for determining the
infinite volume low energy coupling coefficients $g_{A}^{0}$,
$g_{\Delta N}$, and $g_{\Delta\Delta}$ as well as the magnetic
couplings $\mu_0$, $\mu_1$, $\mu_C$, and $\mu_T$. With the above
results for the magnetic moment and previous
results\cite{Smigielski:2007pe} for the axial charge, both
quantities can be coherently analyzed in the $\epsilon$-regime.

\acknowledgments The author would like to thank M. Savage, W.
Detmold, and B. Smigielski for many useful discussions.

\appendix
\section{Zero Mode Functions}
\begin{figure}[!h]
\centering
\includegraphics[scale=1]{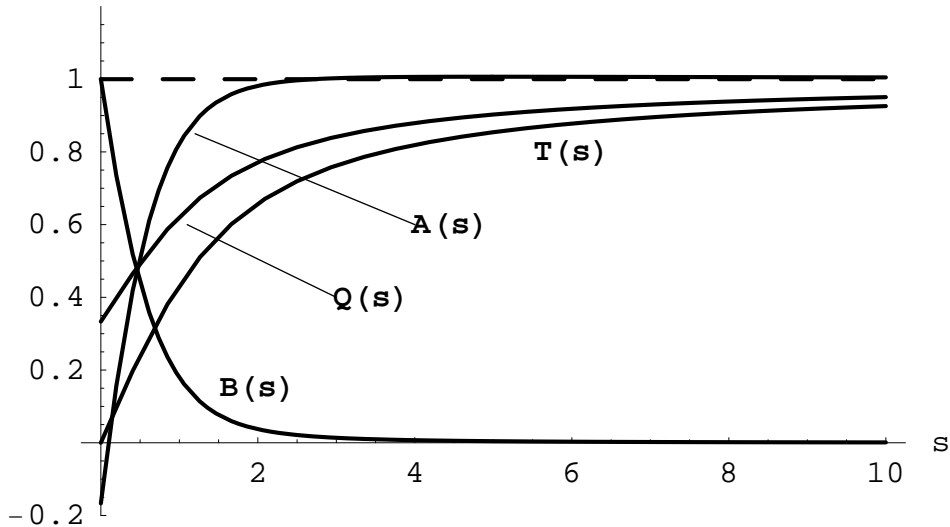}
\caption{The form of the four zero mode functions used in the
calculations.} \label{zerofunct}
\end{figure}

The zero mode functions (with $s=\frac{1}{2}m_{\pi}^2f^2V$)
calculated for the contributing diagrams are:
\begin{eqnarray}
    Q(s)&=&\frac{1}{3}\left(1+2\frac{I_2(2s)}{I_1(2s)}\right)\\
    T(s)&=&\frac{I_{0}(2s)}{I_{1}(2s)}-\frac{1}{s}\\
    A(s)&=&\frac{11}{6}+\frac{1}{3s}+2s-\frac{I_{0}(2s)}{3I_{1}(2s)}-\frac{2s
    I_{0}(2s)}{I_{1}(2s)}\\
    B(s)&=&-\frac{3}{5}-\frac{1}{5s}-\frac{8s}{5}+\frac{I_{0}(2s)}{5I_{1}(2s)}+\frac{8sI_{0}(2s)}{5I_{1}(2s)}.
\end{eqnarray}
Of these functions $Q(s)$ and $T(s)$ have been calculated previously
in refs. \cite{Smigielski:2007pe,Gasser:1987ah}. The functions
$A(s)$ and $B(s)$ are new to this paper and are generated from
contributions of the type shown in figs. (\ref{e3graphs}a) and
(\ref{e3graphs}b). These functions are plotted in fig.
(\ref{zerofunct}).

\section{Sums}
For the calculation of the diagrams given above several sums will be
important. Many of these sums have been calculated previously
\cite{Luscher:1986pf,Edery:2005bx,Detmold:2004ap,Beane:2007qr} and
the divergent sums are defined through dimensional regularization:
\begin{eqnarray}
    c_4&=&\sum_{\vec{n}\not=0}\frac{1}{|\vec{n}|^4}=16.532315\nonumber\\
    c_3&=&\sum_{\vec{n}\not=0}\frac{1}{|\vec{n}|^3}=3.8219235\nonumber\\
    c_2&=&\sum_{\vec{n}\not=0}\frac{1}{|\vec{n}|^2}=-8.913633\nonumber\\
    c_1&=&\sum_{\vec{n}\not=0}\frac{1}{|\vec{n}|}=-2.8372974\nonumber\\
    c_0&=&\sum_{\vec{n}\not=0}1=-1.
\end{eqnarray}
The sum that needs to be evaluated for the two loop diagram is more
complicated, but can be evaluated using the same general techniques:
\begin{eqnarray}
    c'_3&=&\sum_{\vec{m},\vec{n}\not=0}\left(\frac{3}{2}\frac{1}{|\vec{n}|(|\vec{m}|+|\vec{n}|)^2}+\frac{|\vec{m}|}{|\vec{n}|^2(|\vec{m}|+|\vec{n}|)^2}\right)=2.722962.
\end{eqnarray}

\bibliography{masterbibtex}

\end{document}